# Calculation of Matrix Elements
# in the Foldy-Wouthuysen Representation

*V.P. Neznamov\*, A.A.Sadovoy\*\*, A.S.Ul'yanov\*\*\**

*RFNC-VNIIEF, Sarov, Russia*

Abstract

The paper compares the methods used to calculate matrix elements of the operator of radail electron coordinates in an arbitrary order in the Foldy-Wouthuysen representation and with the use of the Dirac equation for 1s-states of the hydrogen-like and helium-like ions of transuranic elements. The obtained analytical and numerical results for 1s-states of the hydrogen-like and helium-like ions proves that the wave function reduction requirement is met with transformation to the Foldy-Wouthuysen representation and confirms that matrix elements can be calculated using only one component (either upper, or lower) of the Dirac bispinor wave function.

*PACS numbers: 03.65.- W, 31.10.+Z.*

---

\* E-mail address:   neznamov@vniief.ru

\*\* E-mail address:   aa_sadovoy@vniief.ru

\*\*\* E-mail address:   A.S.Ulyanov@vniief.ru

1. **INTRODUCTION**

Research into the relativistic system properties is important from viewpoint of theory and applied sciences. The elementary particle physics studied and continues studying various bound systems of leptons and baryons. For applied research, of interest, for example, are the properties of heavy and transuranic ions, where relativistic effects have essential values.

Solution of relativistic equations associates with considerable difficulties. So, the development of methods that allow easier calculation of the relativistic system's properties seems to be an urgent problem.

The paper describes the use of the wave functions in the Foldy-Wouthuysen representation [1], [2] and analytical solutions to Dirac equations for the hydrogen- and helium-like ions of heavy and transuranic elements [3], [4] to demonstrate the Foldy-Wouthuysen representation advantages in calculations of matrix element for various operators, which can be represented in the form of expansion in series in electron coordinates.

## 2. SOME FEATURES OF THE FOLDY-WOUTHUYSEN REPRESENTATION

The Foldy-Wouthuysen (FW) representation was introduced in the paper [1]. The following unitary transformation was used to derive the FW representation from the Dirac representation.

For free motion of particles, the Dirac equation looks like

$$p_0 \psi_D(\vec{x},t) = (H_0)_D \psi_D(\vec{x},t) = (\vec{\alpha}\vec{p} + \beta m)\psi_D(\vec{x},t). \qquad (1)$$

Using the unitary transformation [1]

$$U_0 = \sqrt{\frac{E+m}{2E}}\left(1 + \frac{\beta\vec{\alpha}\vec{p}}{E+m}\right), \quad E = \sqrt{m^2 + \vec{p}^2},$$ we transform Eq. (1) to the form

$$p_0 \psi_{FW}(\vec{x},t) = (H_0)_{FW} \psi_{FW}(\vec{x},t) = \beta E \psi_{FW}(\vec{x},t). \qquad (2)$$

In Eq. (2),

$$(H_0)_{FW} = U_0 (H_0)_D U_0^\dagger = \beta E \; ; \; \psi_{FW}(\vec{x},t) = U_0 \psi_D(\vec{x},t).$$

Eqs. (1), (2) and those given below are written in the system of units $\hbar = c = 1$; the scalar product of four-vectors is taken in the form

$$xy \equiv x^\mu y_\mu = x^0 y^0 - x^\kappa y^\kappa, \quad \mu = 0,1,2,3; \quad \kappa = 1,2,3,$$

$p^\mu = i\left(\dfrac{\partial}{\partial x_\mu}\right)$; $\psi_D(\vec{x},t), \psi_{FW}(\vec{x},t)$ - are four-component wave functions;

$\vec{\alpha} = \beta\vec{\gamma} = \begin{pmatrix} 0 & \vec{\sigma} \\ \vec{\sigma} & 0 \end{pmatrix}, \beta \equiv \gamma^0 = \begin{pmatrix} I & 0 \\ 0 & -I \end{pmatrix}$ are Dirac matrices; $\sigma^i$ are two-component Pauli matrices.

Solutions to Eq. (1) are the wave functions with the positive and negative energy values:

$$\psi_{FW}^{(+)}(x,s) = \frac{1}{(2\pi)^{3/2}} U_s e^{-ipx}, \quad \psi_{FW}^{(-)}(x,s) = \frac{1}{(2\pi)^{3/2}} V_s e^{ipx},$$

$$p_0 = \sqrt{(\vec{p}^2 + m^2)}. \qquad (3)$$

In expression (3), $U_s = \begin{pmatrix} \varphi_s \\ 0 \end{pmatrix}$, $V_s = \begin{pmatrix} 0 \\ \chi_s \end{pmatrix}$, $\varphi_s$ and $\chi_s$ are the two-component normalized Pauli's spin functions. The following orthonormalization and completeness correlations are valid for $U_s$ and $V_s$

$$U_s^\dagger U_{s'} = V_s^\dagger V_{s'} = \delta_{ss'}; \quad U_s^\dagger V_{s'} = V_s^\dagger U_{s'} = 0;$$

$$\sum_s (U_s)_\gamma (U_s^\dagger)_\delta = \frac{1}{2}(1+\beta)_{\gamma\delta}; \qquad (4)$$

$$\sum_s (U_s)_\gamma (U_s^\dagger)_\delta = \frac{1}{2}(1-\beta)_{\gamma\delta}.$$

In expressions (3) and (4) $\gamma$, $\delta$ are spinor indexes, $s$ is spin index. Later, we'll omit the summation symbol and indexes themselves, while summing up with respect to spinor indexes.

In case of interaction between a Dirac particle and external static fields, the closed FW representation exists if only the requirement of commutation of the even and odd parts of the Dirac Hamiltonian is met [5]. By definition, an even operator does not mix the upper and lower components of wave function.

In general case, when fermion interacts with an arbitrary boson field, the problem of transformation from the Dirac representation to the FW representation becomes much more complicated. The general form of the exact FW transformation has been found by Eriksen [6] for arbitrary static external fields. The Eriksen transformation is a one-step transformation of the Dirac wave function and Dirac Hamiltonian to the FW representation.

Another way of direct transformation to the FW representation offered in the paper [7] by one of the given paper authors (see also the review paper [8]) for a general case of interaction with an arbitrary boson field is to derive the relativistic Hamiltonian in the form of a series in terms of the coupling constant powers.

In addition to some direct methods of obtaining Hamiltonians in the FW representation, there are a lot of methods for step-by-step derivation of Hamiltonians free of odd operators. One of them was used, in particular, in the classic Foldy-Wouthuysen work (see [1]) to obtain Hamiltonian in the presence of static external electromagnetic field in the form of a series in terms of $1/m$ powers.

The papers [2], [7], [9] show that step-by-step transformation methods lead to the FW representation only for the first one, or two steps.

The paper [2] studies the main properties of wave functions in the FW representation and establishes a unique relation between the two wave functions in the Dirac and Foldy-Wouthuysen representations.

Diagonalization of Hamiltonian relative to the upper and lower components of wave function is a necessary requirement of transformation from the Dirac representation to FW representation.

The second requirement for the FW transformation is that the upper and lower components of bispinor wave function $\psi_D(\vec{x},t) = A \begin{pmatrix} \varphi(\vec{x},t) \\ \chi(\vec{x},t) \end{pmatrix}$ must be zeroed and the normalization operator of wave function $\psi_D(x)$ must be transformed to the unit operator. The paper [2] proves this requirement and calls it the "wave function reduction requirement".

For the time-independent Dirac Hamiltonian (the case of a free particle, or static external fields), this condition can be represented in the following form:

$$\psi_D^{(+)}(\vec{x},t) = e^{-iEt} A_+ \begin{pmatrix} \varphi^{(+)}(\vec{x}) \\ \chi^{(+)}(\vec{x}) \end{pmatrix} \rightarrow \psi_{FW}^{(+)}(\vec{x},t) = e^{-iEt} \begin{pmatrix} \varphi^{(+)}(\vec{x}) \\ 0 \end{pmatrix};$$

$$\psi_D^{(-)}(\vec{x},t) = e^{iEt} A_- \begin{pmatrix} \varphi^{(-)}(\vec{x}) \\ \chi^{(-)}(\vec{x}) \end{pmatrix} \rightarrow \psi_{FW}^{(-)}(\vec{x},t) = e^{iEt} \begin{pmatrix} 0 \\ \chi^{(-)}(\vec{x}) \end{pmatrix}.$$

(5)

In this equation, $E$ is the particle energy operator's module; $A_+$ and $A_-$ are normalization operators, which, in general, may be different for the positive-energy and negative-energy solutions. Definition of operators $A_+$ and $A_-$ implies that the wave functions $\psi_D^{(+)}(\vec{x},t)$, $\psi_D^{(-)}(\vec{x},t)$ and the spinors $\varphi^{(+)}(\vec{x})$, $\chi^{(-)}(\vec{x})$ are normalized to unity:

$$\int \psi_D^{(\pm)\dagger}(\vec{x},t) \psi_D^{(\pm)}(\vec{x},t) dV = 1, \int \varphi^{(+)\dagger}(\vec{x}) \varphi^{(+)}(\vec{x}) dV = 1, \int \chi^{(-)\dagger}(\vec{x}) \chi^{(-)}(\vec{x}) dV = 1.$$

Pluses and minuses denote states with positive and negative energy, respectively.

For a free particle,

$$E = \sqrt{m^2 + \vec{p}^2}, \quad A_+ = A_- = \sqrt{\frac{E+m}{2E}},$$

(6)

$\varphi^{(+)}(\vec{x}) = e^{i\vec{p}\vec{x}} \varphi$ and $\chi^{(+)}(\vec{x}) = e^{-i\vec{p}\vec{x}} \chi$ for the positive and negative energy solutions, respectively; $\varphi$ and $\chi$ are the two-component Paili's spin functions (see expression (3)).

Functions $\psi_D^{(\pm)}(\vec{x},t)$ and $\psi_{FW}^{(\pm)}(\vec{x},t)$ are the appropriate solutions of the Dirac equation and the equation transformed to the FW representation for a free particle and a particle moving in static external fields. The reduction condition implies transformation of the Dirac wave function to the form $\psi_{FW}^{(\pm)}(\vec{x},t)$ with unit normalization operator.

In general, the Dirac and FW Hamiltonians depend on time. In this case, the reduction condition (5) has the same meaning. When solving specific problems of physics (at least, with the use of the perturbation theory), we use series expansions in the Dirac equation solutions obtained either for free motion of particles, or for particle motion in the presence of static external fields.

In this work in the Dirac and FW representations (with the use of condition (5)) we calculate matrix elements for the momentum operator of electron coordinates of an arbitrary power ($r^n$) using the Dirac wave functions of the hydrogen-like and helium-like ions determined in [3], [4] and [12]. Really, these examples are additional tests for the wave function reduction requirement (5) and they demonstrate an easier way of matrix elements calculating using only one component (either upper, or lower) of the Dirac bispinor wave function.

## 3. MATRIX ELEMENTS OF AN OPERATOR OF ELECTRON COORDINATES IN AN ARBITRARY POWER

### 3.1 Hydrogen-like ions

In general, the wave function the hydrogen-like ion is determined [3] as

$$\Psi(\vec{r}) = \begin{pmatrix} f(r)\Omega_{jlm}(\vec{\Omega}) \\ (-1)^{\frac{1+l-l'}{2}} g(r)\Omega_{jl'm}(\vec{\Omega}) \end{pmatrix}, \qquad (7)$$

where its angular part $\Omega_{jlm}(\vec{\Omega})$ are the 3D spherical spinors, $j$ is the total momentum, $l$ is the orbital momentum, $m$ is the total momentum projection, $r$ is the electron coordinate module.

Normalization

$$\int_0^\infty (f^2 + g^2) r^2 dr = 1 \qquad (8)$$

for state $1s_{1/2}$ gives us functions $f$ and $g$ of the form

$$f = \frac{(2\lambda)^{3/2} \sqrt{E+2}}{\sqrt{2\Gamma(2\gamma+1)}} e^{-\lambda r} (2\lambda r)^{\gamma-1}; \qquad (9)$$

$$g = -\frac{(2\lambda)^{3/2} \sqrt{-E}}{\sqrt{2\Gamma(2\gamma+1)}} e^{-\lambda r} (2\lambda r)^{\gamma-1}, \qquad (10)$$

where $\lambda = \sqrt{-E(E+2)}$, $\gamma = \sqrt{1-(Z\alpha)^2}$, $E$ is the system's binding energy in units of mc², unit of length equals classical radius of electron $\frac{e^2}{mc^2}$, $\alpha = \frac{e^2}{\hbar c}$.

In case of normalization (8), the $n$-th order momentum of electron coordinate is determined by the expression:

$$\langle r^n \rangle = \int_0^\infty (f^2 + g^2) r^{n+2} dr = \frac{1}{(2\lambda)^n} \frac{\Gamma(2\gamma+n+1)}{\Gamma(2\gamma+1)}. \qquad (11)$$

With unit normalization of the wave function's upper component $\tilde{f}(r)$

$$\int_0^\infty \tilde{f}^2(r) r^2 dr = 1, \qquad (12)$$

normalization (8) can be written as

$$A^2 \int_0^\infty \left( \tilde{f}^2(r) + g^2(r) \right) r^2 dr = 1, \qquad (13)$$

where
$$\tilde{f}(r) = \frac{(2\lambda)^{3/2}}{\sqrt{\Gamma(2\gamma+1)}} \cdot e^{-\lambda r} (2\lambda r)^{\gamma-1} \qquad (14)$$

and $A = \sqrt{\dfrac{2-E}{2}}$.

It is clear that in this case expression (11) for momentum $\langle r^n \rangle$ remains unchanged.

According to condition (5), the wave function in FW representation is for our case the wave function's upper component (14) normalized to unity.

One may calculate momentum of electron coordinate with normalization (12) and with the use of the wave function's upper component alone.

$$\langle r^n \rangle = \int_0^\infty \tilde{f}^2 r^{n+2} dr = \frac{(2\lambda)^3}{\Gamma(2\gamma+1)} \int e^{-2\lambda r} (2\lambda r)^{\gamma-1} r^{n+2} dr = \frac{1}{(2\lambda)^n} \frac{\Gamma(2\gamma+n+1)}{\Gamma(2\gamma+1)}. \qquad (15)$$

One can see that expression (15) coincides with expression (11).

The absolute values of moment of electron coordinates for transuranic elements with $n = -2,...,3$ are given in Table 1.

**Table 1 – Radial moment of the hydrogen-like ions of transuranic elements in state $1s_{1/2}$.**

| Name of element | Nuclear charge Z | $\langle r^{-2} \rangle$ | $\langle r^{-1} \rangle$ | $\langle r^1 \rangle$ | $\langle r^2 \rangle$ | $\langle r^3 \rangle$ |
|---|---|---|---|---|---|---|
| U | 92 | 2.522 | 0.906 | 1.849 | 4.795 | 16.009 |
| Np | 93 | 2.674 | 0.924 | 1.819 | 4.650 | 15.310 |
| Pu | 94 | 2.840 | 0.943 | 1.790 | 4.508 | 14.642 |
| Am | 95 | 3.021 | 0.962 | 1.761 | 4.371 | 14.003 |
| Cm | 96 | 3.219 | 0.982 | 1.733 | 4.238 | 13.393 |
| Bk | 97 | 3.437 | 1.002 | 1.704 | 4.109 | 12.809 |
| Cf | 98 | 3.677 | 1.023 | 1.677 | 3.984 | 12.25 |
| Es | 99 | 3.942 | 1.045 | 1.649 | 3.862 | 11.715 |
| Fm | 100 | 4.238 | 1.067 | 1.622 | 3.743 | 11.203 |
| Mv | 101 | 4.570 | 1.090 | 1.596 | 3.628 | 10.712 |

### 3.2 Helium-like ions

For the helium-like ions, we use the Dirac equation solution with minimum approximation by the method of multidimensional angular Coulomb functions [13]. The wave function of state $0^+$ of the helium-like ions corresponding to configuration $1s^2$ has the form

$$\Psi(\vec{r}_1,\vec{r}_2) = \begin{pmatrix} \Phi(\vec{r}_1,\vec{r}_2) \\ X(\vec{r}_1,\vec{r}_2) \end{pmatrix} = \frac{1}{\rho^{\frac{5}{2}}} \begin{pmatrix} M(\rho)U(\vec{\Omega}_1,\vec{\Omega}_2) \\ N(\rho)W(\vec{\Omega}_1,\vec{\Omega}_2) \end{pmatrix}, \qquad (16)$$

where $\vec{\rho} = \vec{r}_1 + \vec{r}_2$ is a collective variable and multidimensional angular functions are Slater determinants

$$U(\vec{\Omega}_1,\vec{\Omega}_2) = \sqrt{\frac{\Gamma(6)}{2!}} \begin{pmatrix} \varphi_+(\vec{\Omega}_1) & \varphi_+(\vec{\Omega}_2) \\ \varphi_-(\vec{\Omega}_1) & \varphi_-(\vec{\Omega}_2) \end{pmatrix}; \qquad W = \sqrt{\frac{\Gamma(6)}{2!}} \begin{pmatrix} \chi_+(\vec{\Omega}_1) & \chi_+(\vec{\Omega}_2) \\ \chi_-(\vec{\Omega}_1) & \chi_-(\vec{\Omega}_2) \end{pmatrix} \qquad (17)$$

from the basic functions

$$\varphi_{\frac{1}{2}0m}(\vec{\Omega}_i) = \sqrt{\frac{1}{\Gamma(3)}} \Omega_{\frac{1}{2}0m}(\vec{\Omega}_i), \quad \chi_{\frac{1}{2}1m}(\vec{\Omega}_i) = \sqrt{\frac{1}{\Gamma(3)}} \Omega_{\frac{1}{2}1m}(\vec{\Omega}_i), \qquad (18)$$

respectively, where $\Omega_{jlm}(\vec{\Omega}_i)$ are the three-dimensional spherical spinors.

The magnetic quantum number is $m = \pm 1/2$, the phase factor of the lower component equals $(-1)^{l-j+\frac{1}{2}} = (-1)^{\frac{-1+0+1}{2}} = 1$, the spinor indexes correspond to the quantum numbers $\{j,l,j_z\}$. The upper and lower spinors are homogeneous polynomials of one and the same power, $K$.

The equation system for the helium-like ions in state $O^+$ has the form [4]

$$\left(E\alpha + \sum_{i=1}^{2} \frac{\alpha Z}{r_i} - \frac{\alpha}{|\vec{r}_1 - \vec{r}_2|}\right)\Phi = -i\sum_{i=1}^{2} \vec{\sigma}_i \vec{p}_i X; \qquad (19)$$

$$\left((E+4)\alpha + \sum_{i=1}^{2} \frac{\alpha Z}{r_i} - \frac{\alpha}{|\vec{r}_1 - \vec{r}_2|}\right)X = -i\sum_{i=1}^{2} \vec{\sigma}_i \vec{p}_i \Phi, \qquad (20)$$

where $\varepsilon$ is the total energy of system, in units of mc$^2$; $E = \varepsilon - A < 0$ is the binding energy of system; unit of length equals classical radius of electron $\frac{e^2}{mc^2}$, Multiplying Eqs. (19) and (20) by the complex-conjugate row $(U^* \ W^*)$ and integrating with respect to each angular variables give us the following expressions for amplitudes of wave function (16) expansion in series in terms of two-components angular functions

$$M' - \frac{5}{2}\frac{1}{\rho}M - \frac{\alpha}{2}\left(E + 4 + \frac{5}{\rho}\left(Z - \frac{5}{16}\right)\right)N = 0; \qquad (21)$$

$$N' + \frac{5}{2}\frac{1}{\rho}N + \frac{\alpha}{2}\left(E + \frac{5}{\rho}\left(Z - \frac{5}{16}\right)\right)M = 0. \qquad (22)$$

The process of solving Eqs. (21) – (22) described in details in the paper [4] similar to that described in the paper [3] for the hydrogen-like ions resulted in the equation for the binding energy of the helium-like ions

$$E = 2\left(\sqrt{1-\alpha^2\left(Z-\frac{5}{16}\right)^2} - 1\right). \tag{23}$$

The binding energy values for some helium-like ions $0^+$ of transuranic elements calculated from Eq. (23) are given in the Table 2.

**Table 2 – The binding energy of the helium-like ions in state $0^+$ of transuranic elements.**

|                 |                      | Binding energy $E, keV$ |         |         |
|-----------------|----------------------|-------------------------|---------|---------|
|                 | Quantum number, $n_r$ | 0                       | 1       | 2       |
| Name of element | Nuclear charge Z     | Helium-like ions        |         |         |
| U               | 92                   | -262.23                 | -249.69 | -134.42 |
| Np              | 93                   | -269.01                 | -254.74 | -137.43 |
| Pu              | 94                   | -275.92                 | -259.84 | -140.48 |
| Am              | 95                   | -282.97                 | -264.99 | -143.58 |
| Cm              | 96                   | -290.17                 | -270.18 | -146.71 |
| Bk              | 97                   | -297.51                 | -275.44 | -149.89 |
| Cf              | 98                   | -305.01                 | -280.74 | -153.11 |
| Es              | 99                   | -312.67                 | -286.10 | -156.37 |
| Fm              | 100                  | -320.48                 | -291.52 | -159.68 |
| Mv              | 101                  | -328.47                 | -296.99 | -163.03 |

With regard to normalization

$$\int_0^\infty \left(M^2(\rho) + N^2(\rho)\right) d\rho = 1, \tag{24}$$

amplitudes of expansion in series of the upper and lower components of the wave function for the main ($n_r = 0$) state $0^+$ of the helium-like ion of transuranic element, which are solutions to the system of Eqs. (21) – (22), have the form

$$M(\rho) = \frac{\sqrt{\lambda}\sqrt{E+4}}{2\sqrt{\Gamma(2\gamma-1)}} \cdot e^{-\frac{\lambda\rho}{2}} (\lambda\rho)^{\gamma-1}; \tag{25}$$

$$N(\rho) = \frac{-\sqrt{\lambda}\sqrt{-E}}{2\sqrt{\Gamma(2\gamma-1)}} \cdot e^{-\frac{\lambda\rho}{2}} (\lambda\rho)^{\gamma-1}, \tag{26}$$

where $\lambda = \sqrt{-E(E+4)}$.

It should be noted that the expansion amplitudes above are proportional (the proportionality constant is $\sqrt{\dfrac{E+4}{-E}}$ ).

The $n$-th order momentum for the helium-like ions in the main state $0^+$ is calculated as

$$\langle r^n \rangle = \langle r^n \rangle_1 + \langle r^n \rangle_2 = \int \frac{1}{\rho^5} MU \sum_{i=1}^{2} r_i^n MU d\vec{r}_1 d\vec{r}_2 + \int \frac{1}{\rho^5} NW \sum_{i=1}^{2} r_i^n NW d\vec{r}_1 d\vec{r}_2. \tag{27}$$

Integration with respect to angular variables for the first term of the $n$-th order momentum gives us

$$\langle r^n \rangle_1 = \frac{\Gamma(6)\Gamma(n+3)}{\Gamma(3)\Gamma(n+6)} \int_0^\infty M^2(\rho) \rho^n d\rho . \tag{28}$$

Similar to (28), we obtain

$$\langle r^n \rangle_2 = \frac{\Gamma(6)\Gamma(n+3)}{\Gamma(3)\Gamma(n+6)} \int_0^\infty N^2(\rho) \rho^n d\rho . \tag{29}$$

As a result, we obtain the following expression for the electron radius momentum of the helium-like ions

$$\langle r^n \rangle = \frac{\Gamma(6)\Gamma(n+3)}{\Gamma(3)\Gamma(n+6)} \int_0^\infty \left( M^2(\rho) + N^2(\rho) \right) \rho^n d\rho . \tag{30}$$

Use amplitudes (25), (26) and obtain from expression (30) the analytical expression for the $n$-th order momentum

$$\langle r^n \rangle = \frac{1}{\lambda^n} \frac{\Gamma(6)\Gamma(n+3)}{\Gamma(3)\Gamma(n+6)} \frac{\Gamma(2\gamma+n-1)}{\Gamma(2\gamma-1)} . \tag{31}$$

With unit normalization of the wave function's upper component $\tilde{M}(r)$

$$\int_0^\infty \tilde{M}^2(\rho) d\rho = 1, \tag{32}$$

normalization (24) can be written in the form

$$A^2 \int_0^\infty \left( \tilde{M}^2(\rho) + N^2(\rho) \right) d\rho = 1, \tag{33}$$

where $A = \sqrt{\dfrac{4-E}{4}}$ and

$$\tilde{M}(\rho) = \frac{\sqrt{\lambda}}{\sqrt{\Gamma(2\gamma-1)}} \cdot e^{-\frac{\lambda\rho}{2}} (\lambda\rho)^{\gamma-1} . \tag{34}$$

Apparently, expression (31) for momentum $\langle r^n \rangle$ remains unchanged.

Similarity to expression (15) momentums $\langle r^n \rangle$ are calculated with the use of the upper component (34) of bispinor wave function (16) normalized to unity.

The electron coordinate moment with normalization (32) are

$$\langle r^n \rangle = \frac{\Gamma(6)\Gamma(n+3)}{\Gamma(3)\Gamma(n+6)} \int_0^\infty \tilde{M}^2(\rho)\rho^n d\rho = \frac{1}{\lambda^n} \frac{\Gamma(6)\Gamma(n+3)}{\Gamma(3)\Gamma(n+6)} \frac{\Gamma(2\gamma+n-1)}{\Gamma(2\gamma-1)} \quad (35)$$

and expression (35) coincides with expression (31) like in case of hydrogen-like ions (see expression (15) and (11)).

The absolute values of moment of various orders for heavy elements (from uranium to mendelevium) in state $0^+$ are given in the Table 3.

Table 3 – Radial moment of the heavy elements' helium-like ions in state $0^+$

| Name of element | Nuclear charge Z | $\langle r^{-2} \rangle$ | $\langle r^{-1} \rangle$ | $\langle r^1 \rangle$ | $\langle r^2 \rangle$ | $\langle r^3 \rangle$ |
|---|---|---|---|---|---|---|
| U  | 92  | 1.774 | 0.900 | 1.762 | 4.302 | 13.495 |
| Np | 93  | 1.852 | 0.918 | 1.731 | 4.156 | 12.283 |
| Pu | 94  | 1.934 | 0.937 | 1.700 | 4.014 | 12.200 |
| Am | 95  | 2.021 | 0.956 | 1.670 | 3.877 | 11.597 |
| Cm | 96  | 2.112 | 0.975 | 1.640 | 3.743 | 11.023 |
| Bk | 97  | 2.209 | 0.996 | 1.610 | 3.614 | 10.474 |
| Cf | 98  | 2.312 | 1.016 | 1.581 | 3.489 | 9.951 |
| Es | 99  | 2.422 | 1.038 | 1.552 | 3.367 | 9.452 |
| Fm | 100 | 2.538 | 1.060 | 1.523 | 3.248 | 8.974 |
| Mv | 101 | 2.662 | 1.083 | 1.494 | 3.133 | 8.518 |

## 4. CONCLUSION

The paper compares the methods used to calculate matrix elements of the operator of radail electron coordinates in an arbitrary order in the Foldy-Wouthuysen representation and with the use of the Dirac equation for 1s-states of the hydrogen-like and helium-like ions of transuranic elements.

By means of the direct calculations we obtain that one and the same operator $r^n$ can be used as an operator of coordinate in Dirac and FW representation for $1s$ - states of hydrogen – like and helium – like ions.

As is known, in general case, these operators are different in these two representations. If the operator $r^n_{FW}$ is used in FW representation, then in Dirac representation it is $r^n_D = U^+_{FW} r^n_{FW} U_{FW}$, where $U_{FW}$ is transformation matrix with a complex dependence on the external fields. In the free case $\vec{r}_D$ is an operator of Newton – Wigner [14].

The obtained analytical and numerical results for 1s-states of the hydrogen-like and helium-like ions proves that the wave function reduction requirement is met with transformation to the Foldy-Wouthuysen representation and confirms that matrix elements can be calculated using only one component (either upper, or lower) of the Dirac bispinor wave function.